\documentclass[aps,pra,showpacs,twocolumn]{revtex4-1}
\usepackage{}
\usepackage{amsmath}
\usepackage{amssymb}
\usepackage{graphicx}
\usepackage{epstopdf}
\usepackage{times,txfonts}
\usepackage{easybmat}
\usepackage{mathtools}
\newcommand{\ket}[1]{|#1\rangle}
\newcommand{\bra}[1]{\langle #1|}
\usepackage[bookmarks=false]{hyperref}
\hypersetup{colorlinks=true,citecolor=blue,linkcolor=blue,urlcolor=blue,pdfstartview=FitH,bookmarksopen=true}
\usepackage{color,soul}

\begin{document}
\title{Coherence-protected nonadiabatic geometric quantum computation}
\author{K. Z. Li}
\affiliation{Department of Physics, Shandong University, Jinan 250100, China}
\author{G. F. Xu}
\email{xgf@sdu.edu.cn}
\affiliation{Department of Physics, Shandong University, Jinan 250100, China}
\author{D. M. Tong}
\email{tdm@sdu.edu.cn}
\affiliation{Department of Physics, Shandong University, Jinan 250100, China}
\date{\today}
\begin{abstract}
Because of using geometric phases, nonadiabatic geometric gates have the robustness against control errors. On the other hand, decoherence still affects nonadiabatic geometric gates, which is a key factor in reducing their fidelities. In this paper, we show that based on the system Hamiltonian that realizes a nonadiabatic geometric gate, one may construct a new system Hamiltonian, by using which not only the geometric feature of the nonadiabatic geometric gate is preserved, but also the system's coherence is protected. As a result, a coherence-protected nonadiabatic geometric gate is realized with the new system Hamiltonian and this gate has the robustness against both control errors and decoherence. We further implement our scheme with nitrogen-vacancy centers and show that a universal set of coherence-protected nonadiabatic geometric gates can be realized. Our scheme does not need auxiliary systems or the encoding of logical qubits with physical qubits, which saves resources for the implementation. Due to the robustness against both control errors and decoherence, our scheme provides a promising way to realize high-fidelity quantum gates.
\end{abstract}
\maketitle

\section{Introduction}
Quantum computation is built by using quantum mechanical principles and therefore has many differences from classical computation. Quantum computation encodes information into superposition states, while classical computation into discrete states. Moreover, it uses quantum logic totally different from the Boolean logic on which classical computation is built. Due to quantum parallelism, quantum computation is believed to be superior to its classical counterpart and examples have been found to support this. For instance, quantum computation has been shown to solve certain problems, such as factoring large integers \cite{Shor} and searching unsorted data \cite{Grover}, faster than classical computation. In the circuit model of quantum computation, information is processed by various quantum gates. Thus, realizing a universal set of high-fidelity quantum gates becomes one key step to using circuit-based quantum computation in practice. However, realizing high-fidelity quantum gates is still a hard work due to noises and efforts in both theory and experiment are needed.

To lower the difficulty of realizing high-fidelity quantum gates, error-resilient quantum gates are proposed and among them, geometric gates play an important role. Because of using geometric phases, geometric gates have the robustness against control errors \cite{Chiara,Solinas2004,Lupo,Filipp,Johansson,Berger,Zhu2005}. Geometric gates \cite{Jones} were first built by using adiabatic Abelian geometric phases \cite{Berry}. Soon, the extension from adiabatic geometric gates to adiabatic holonomic gates \cite{Duan,Zanardi}, which are based on adiabatic non-Abelian geometric phases \cite{Wilczek}, was implemented. The common feature of adiabatic geometric gates and adiabatic holonomic gates is the requirement of adiabatic evolutions. Then to relax such a requirement, nonadiabatic geometric gates \cite{WangXB,Zhu}, which are based on Aharonov-Anandan phases \cite{Aharonov}, and nonadiabatic holonomic gates \cite{Sjoqvist2012,Xu2012}, which are based on nonadiabatic non-Abelian geometric phases \cite{Anandan}, were proposed. Since the requirement of adiabatic evolutions is removed, nonadiabatic geometric gates can be realized with high-speed implementations. Due to the merits of geometric robustness and high-speed implementations, nonadiabatic geometric gates have been attracting much attention \cite{WANG2001,Zhu2003,Zheng,Sjoqvist2003,Chen2006,Cen2006,Feng2007,Kim,Wu2007,
Feng2009,Chen2012,Zhao2016,Zhu2003PRA,Ota,Zhang2005,Solinas,Oto2009, Zhu2016,Thomas2011,Xu2014PRA,Xu2014SR,Zhao,Xue,Liu,Du,Leibfried,Yin,Sun,guo2018,xia2020,su2020,own2020,xue2020,npj2018}. Moreover, they have been experimentally demonstrated with various physical systems, such as trapped ions \cite{Leibfried}, nuclear magnetic resonance \cite{Du}, nitrogen-vacancy (NV) centers in diamond \cite{npj2018} and superconducting circuits \cite{guo2018,Sun,Yin}.

While nonadiabatic geometric gates have the robustness against control errors, decoherence can still affect them and this is a key factor in reducing their fidelities. Thus, an important and necessary topic is to strengthen the robustness of nonadiabatic geometric gates and make them also robust against decoherence. In this paper, we propose a scheme to realize coherence-protected nonadiabatic geometric gates, which have robustness against control errors and decoherence. We first demonstrate a general way to construct the system Hamiltonian $H_S$ that generates the desired nonadiabatic geometric gate. Then we take decoherence into account and show that based on $H_S$ one may construct a new system Hamiltonian $H^\prime_S$ with which coherence-protected nonadiabatic geometric gates can be realized. We further demonstrate the specific realization procedure of a universal set of coherence-protected nonadiabatic geometric gates with NV centers. Our scheme is a system Hamiltonian designing scheme and does not need auxiliary systems or encoding. This makes our scheme easy to implement and advantageous over previous schemes. Thus our scheme is helpful to realize more efficient and robust quantum gates.

\section{coherence-protected nonadiabatic geometric gates}

We now demonstrate our scheme. Consider a physical system with an $N$-dimensional Hilbert space and our aim is to realize nonadiabatic geometric gates acting on this physical system. To this end, we consider $N$ orthonormal states $\ket{\nu_k(t)}$ and they satisfy $\ket{\nu_{k}(T)}=\ket{\nu_{k}(0)}$, where $k=1,\ldots,N$ and $T$ is the total evolution time. Based on $\ket{\nu_k(t)}$, one can construct another $N$ orthonormal states $\ket{\phi_k(t)}=e^{i\gamma_k(t)}\ket{\nu_k(t)}$, where $\gamma_k(t)=i\int^{t}_{0}\langle\nu_k(t^{\prime})\ket{\dot{\nu}_k(t^{\prime})}dt^{\prime}$.
Substituting the states $\ket{\phi_k(t)}$ into $H_S=i\sum^{N}_{k=1}\ket{\dot{\phi}_k(t)}\bra{\phi_k(t)}$, we immediately have
\begin{align}
H_S=i\sum_{l\neq k}^{N}\langle\nu_{l}(t)\ket{\dot{\nu}_{k}(t)}\ket{\nu_{l}(t)}\bra{\nu_{k}(t)}.
\label{hs}
\end{align}
One can find that if $H_S$ is considered to be the system Hamiltonian of the $N$-dimensional physical system, a nonadiabatic geometric gate acting on this physical system can be realized. To see this, we consider the $N$-dimensional physical system is initially in state $\ket{\phi_k(0)}$. Then the physical system will evolve along the path described by $\ket{\phi_k(t)}=\mathcal {T}e^{-i\int_0^t H_{S}dt^\prime}\ket{\phi_k(0)}$, and to the final state $\ket{\phi_k(T)}=e^{i\gamma_k(T)}\ket{\phi_k(0)}$ at the end of the evolution. Based on this, the evolution operator generated by the system Hamiltonian $H_S$ can be written as
\begin{align}
U(T)=\sum_{k}e^{i\gamma_k(T)}\ket{\phi_{k}(0)}\bra{\phi_{k}(0)}, \label{UT}
\end{align}
where $\gamma_k(T)=
\arg\langle\nu_k(0)\ket{\nu_k(T)}+i\int^{T}_{0}\langle\nu_k(t)
\ket{\dot{\nu}_k(t)}dt$. It is noteworthy that the first term in $\gamma_k(T)$ equals to zero because of $\ket{\nu_{k}(T)}=\ket{\nu_{k}(0)}$ and we keep this term when emphasizing the $U(1)$ gauge invariance of nonadiabatic Abelian geometric phases. It can be verified that the dynamical phases keep zero during the evolution, i.e.,
\begin{align}
\bra{\phi_k(t)}H_S\ket{\phi_k(t)}=\bra{\nu_k(t)}H_S\ket{\nu_k(t)}=0.
\end{align}
Thus, the phases $\gamma_k(T)$ are geometric phases. As a result, the gate $U(T)$ is a nonadiabatic geometric gate.

While the nonadiabatic geometric gate $U(T)$ realized by $H_S$ has the geometric robustness, it can be affected by decoherence, which is a key factor in reducing its fidelity. We next demonstrate that based on the system Hamiltonian $H_S$, one may construct a new system Hamiltonian $H^\prime_S$ and by using it, one can not only preserve the geometric feature of the nonadiabatic geometric gate $U(T)$, but also protect the system's coherence. We construct the new system Hamiltonian $H^\prime_S$ inspired by continuous dynamical decoupling. Continuous dynamical decoupling provides a promising way to suppress decoherence and it can be designed to multitask: suppressing decoherence and realizing quantum gate can be implemented at the same time \cite{udd1,udd2,udd3,udd4,udd5,udd6}. To construct $H^\prime_S$, we consider the evolutions of the $N$-dimensional system and its environment together. Without loss of generality, the Hamiltonian of the $N$-dimensional system and its environment can be written as
\begin{align}
H_{\text{tot}}=H^\prime_{S}+H_{E}+H_{SE},
\end{align}
where $H^\prime_{S}$ is the new system Hamiltonian, $H_{E}$ is the environment Hamiltonian, and $H_{SE}$ is the interaction Hamiltonian between the system and its environment. Here, we consider the case that the environment Hamiltonian $H_E$ and the interaction Hamiltonian $H_{SE}$ is fixed and given. To realize coherence-protected nonadiabatic geometric gates, we suppose the system Hamiltonian $H^\prime_{S}$ has the following form
\begin{align}
H^\prime_{S}=V(t)H_SV^{\dag}(t)+i\frac{dV(t)}{dt}V^{\dag}(t). \label{HS}
\end{align}
In the above, $V(t)$ is a unitary operator that satisfies the following two conditions. Condition one: $V(t)$ is periodic with respect to time $t$ and turns into the identity operator $I_N$ at the end of each time period,
\begin{align}
V(t+n\tau) = V(t),~~~V(n\tau)=I_N, \label{condition1}
\end{align}
where $n$ is a positive integer and $\tau$ is the minimal positive period of $V(t)$. The period $\tau$ is supposed to be much smaller than the environment correlation time $\tau_{e}$, and then the time dependence of $H_{SE}$ over the timescale $\tau$ can be neglected. Condition two: the unitary operator $V(t)$ needs to satisfy
\begin{align}
\int^{\tau}_{0}V^{\dag}(t)H_{SE}V(t)dt=0, \label{condition2}
\end{align}
where the interaction Hamiltonian $H_{SE}$ is given. The conditions in Eqs.~(\ref{condition1}) and (\ref{condition2}) are usually used in the field of continuous dynamical decoupling and many previous works have shown they can be satisfied in practice. In the following, we show that with the system Hamiltonian $H^\prime_S$, one can not only preserve the geometric feature of the nonadiabatic geometric gate $U(T)$, but also protect the system's coherence.

We first show the preservation of the geometric feature of the nonadiabatic geometric gate $U(T)$. We consider the evolution driven by the Hamiltonian $H^\prime_S$ alone. Suppose the $N$-dimensional physical system is initially in the state $\ket{\phi_k(0)}$ with $k\in\{1,\cdots,N\}$. Then the system will evolve along the path described by $\ket{\tilde{\phi}_k(t)}=e^{i\gamma_{k}(t)}V(t)\ket{\nu_{k}(t)}$ and to the final state $\ket{\tilde{\phi}_k(T)}=e^{i\gamma_k(T)}\ket{\phi_k(0)}$ at the end of the evolution. By calculation, the phase accumulated during the cyclic evolution reads
\begin{align}
\gamma_k(T)&=\int^{T}_{0} i \langle\tilde{\nu}_k(t)\ket{\dot{\tilde{\nu}}_k(t)}
-\bra{\tilde{\nu}_k(t)}H^\prime_{S}
\ket{\tilde{\nu}_k(t)}dt \nonumber \\
&=\arg\langle\nu_k(0)\ket{\nu_k(T)}+i\int^{T}_{0}\langle\nu_k(t)
\ket{\dot{\nu}_k(t)}dt,
\end{align}
where $\ket{\tilde{\nu}_{k}(t)}=V(t)\ket{\nu_{k}(t)}$. One can see that the system evolves cyclically if its initial state is $\ket{\phi_k(0)}$. Moreover, the phases $\gamma_k(T)$ in the above equation are same as that in Eq. (\ref{UT}). Therefore, $\gamma_k(T)$ are geometric phases and the geometric feature of the nonadiabatic geometric gate $U(T)$ is preserved. We second show that with the system Hamiltonian $H^\prime_S$, one can also protect the system's coherence. For convenience, we move to the rotating frame defined by $V(t)$. In this rotating frame, the total Hamiltonian takes the form
\begin{align}\label{eq.8}
H_{r}&=V^{\dag}(t)\Big[(H^\prime_{S}+H_{SE}+H_{E})-i\frac{dV(t)}{dt}V^{\dag}(t)\Big]V(t)
\nonumber \\
&=H_S+H_{E}+V^{\dag}(t)H_{SE}V(t).
\end{align}
Generally, $H_S$ in the above equation is time-dependent. But by dividing the whole evolution process into several segments, one can obtain a piecewise $H_S$ which is time-independent in each segment and we here consider this case. We use $H_{j}$ to denote the value of $H_S$ in the $j$-th segment of the evolution, i.e., $H_S=H_{j}$ when $t_{j}\leq{t}\leq{t}^\prime_{j}$, where $t_{j}$ and $t^\prime_{j}$ are respectively the initial time and end time of the $j$-th segment. Moreover, we require the time duration of each segment is an integral multiple of $\tau$, i.e., $t^\prime_{j}-t_{j}=m_{j}\tau$ with $m_j$ being a positive integer. As illustrated before, the total evolution time $T=M\tau$, with $M$ being a positive integer. Thus, the relation between $m_j$ and $M$ is $\sum_jm_j=M$. We use $m$ to denote a positive integer which satisfies $0\leq{m}\leq{M}$. Then time $t=m\tau$ must belong to some segment. Without loss of generality, we suppose time $t=m\tau$ belongs to the $j$-th segment, i.e., $t_{j}\leq{t=m\tau}\leq{t^\prime_{j}}$. Then the evolution operator generated by the Hamiltonian $H_r$ at time $t=m\tau$ takes the form
\begin{align}
U_{r}(t)&=\mathcal{T}\exp\Big(-i\int^{m\tau}_{t_{j}}H_{r,j}dt^\prime\Big)U_{r}
(t^\prime_{j-1})
\nonumber \\
&=\Big[\mathcal{T}\exp\Big(-i\int^{t_{j}+\tau}_{t_{j}}H_{r,j}dt^\prime\Big)
\Big]^{(m\tau-t_j)/\tau}U_{r}(t^\prime_{j-1}), \label{urt}
\end{align}
where $H_{r,j}=H_j+H_{E}+V^{\dag}(t)H_{SE}V(t)$ and $U_{r}
(t^\prime_{j-1})$ is the evolution operator generated by $H_r$ at time $t^\prime_{j-1}$. By using the Magnus expansion, one can have
\begin{align}
\mathcal{T}\exp\Big(-i\int^{t_{j}+\tau}_{t_{j}}H_{r,j}dt^\prime\Big)
=e^{-i(\bar{H}^{(0)}+\bar{H}^{(1)}+\ldots)\tau},
\end{align}
where the term $\bar{H}^{(j)}$ is an operator of order $\tau^{j}$. We consider only the first order term $\bar{H}^{(0)}=\frac{1}{\tau}\int^{t_{j}+\tau}_{t_{j}}H_{r,j}dt^\prime$, which is known as a lowest-order high-frequency approximation. This approximation is exact in the limit $\tau\rightarrow0$. In a realistic scenario, we require that $\tau$ is much smaller than the enviroment correlation time $\tau_{e}$, that is $\tau/\tau_{e}\ll1$. In the limit $\tau\rightarrow0$ while keeping $t^\prime_{j}-t_{j}=m_{j}\tau$ constant, the evolution operator in Eq.~(\ref{urt}) can be expressed as
\begin{align}
U_{r}(t)&=(e^{-i\bar{H}^{(0)}\tau})^{(m\tau-t_j)/\tau}U_{r}(t^\prime_{j-1}) \nonumber \\
&=[e^{-iH_{j}(t-t_{j})}\otimes{e^{-iH_{E}(t-t_{j})}}]U_{r}(t^\prime_{j-1}).
\end{align}
Moving back to the original frame, one can get the evolution operator generated by $H_{\text{tot}}$,
\begin{align}
U_{\text{tot}}(t)= [V(t)e^{-iH_{j}(t-t_{j})}\otimes{e^{-iH_{E}(t-t_{j})}}]U_{r}(t^\prime_{j-1}).
\end{align}
Then, the evolution operator generated by $H_{\text{tot}}$ at the end time $T$ reads
\begin{align}
U_{\text{tot}}(T)=\Big(\prod_{j} e^{-iH_{j}m_{j}\tau}\Big)\otimes{U_E}=U(T)\otimes{U_E},
\end{align}
where $U(T)$ and $U_E$ are the evolution operators of the system and  environment, respectively. The above equation shows that with the system Hamiltonian $H^\prime_S$, one can decouple the system from its environment and protect the system's coherence. We have shown that the gate $U(T)$ is a nonadiabatic geometric gate. Thus, the system Hamiltonian $H^\prime_S$ generates a coherence-protected nonadiabatic geometric gate, which has the robustness against both control errors and decoherence.

\section{Implementing universal gates}

In the above section, we have generally discussed the construction of the new system Hamiltonian $H^\prime_S$ and shown that by using it, one can preserve the geometric feature of the nonadiabatic geometric gate and meanwhile protect the system's coherence. Based on these results, we here implement our scheme with NV centers. NV centers have attracted much attention due to fast resonant spin manipulation, easy initialization and readout by laser, and the potential to operate at room temperature. We in the following show that a universal set of coherence-protected nonadiabatic geometric gates can be realized with such systems.

\subsection{One-qubit gates}

\begin{figure}[t]
\begin{center}
\includegraphics[scale=0.275]{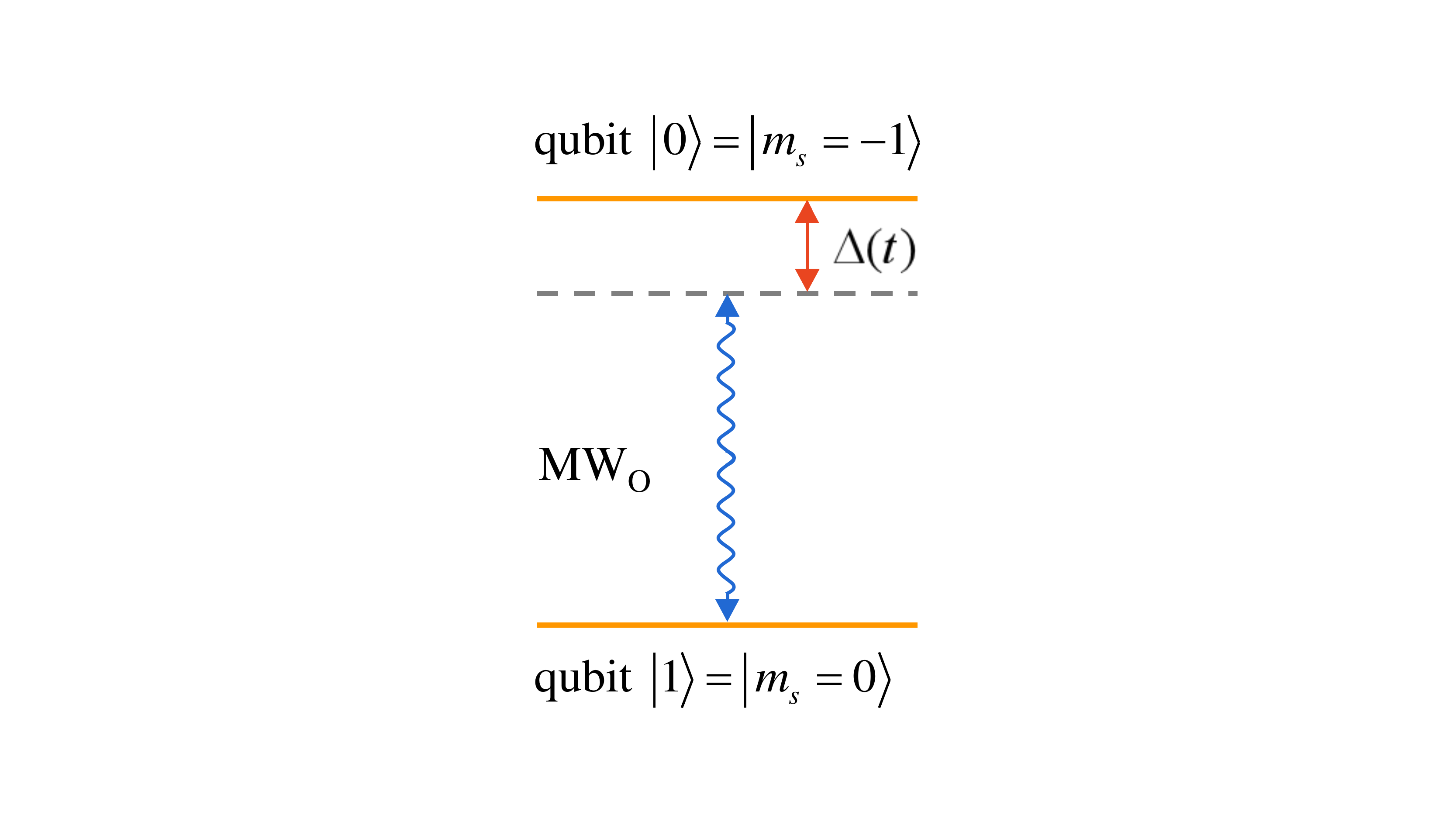}
\end{center}
\caption{(Color online) The level structure for one-qubit gates. Zeeman levels $\ket{m_{s}=-1}$ and $\ket{m_{s}=0}$ are used to encode the logical states $\ket{0}$ and $\ket{1}$. The wavy arrow line indicates the level-selective coupling of microwave field MV$_{O}$, with adjustable Rabi frequency, phase and detuning.
}\label{Fig1}
\end{figure}

We first demonstrate the realization of the one-qubit gates. Consider a NV center in diamond with a proximal $^{13}\text{C}$ atom. This system has a spin-triplet ground state and the nearby nuclear spins ($^{13}\text{C}$ and the host $^{15}\text{N}$) are polarized \cite{ref10}. We take two Zeeman levels $\ket{m_{s}=-1}\equiv\ket{0}$ and $\ket{m_{s}=0}\equiv\ket{1}$ as the qubit basis states, as shown in Fig.~\ref{Fig1}.

The loss of quantum coherence of such a NV center in the high-purity type IIa diamond is principally caused by the hyperfine interaction with the surrounding $^{13}\text{C}$ nuclear spin bath \cite{nv1,nv2,nv3}, which can be described by a random local magnetic field (the Overhauser field). The dynamical fluctuation of the local Overhauser field driven by pairwise nuclear-spin flip flop are much slower than the typical gate time, therefore the local Overhauser field can be seen as a random time-independent variable \cite{nv3,nv4,nv5}. Due to the linear dependence of the energies $\ket{m_{s}=\pm1}$ on the magnetic field, the impact of the spin bath on the NV center can be described by a pure-dephasing model and the interaction Hamiltonian can be written as
\begin{align}
H_{SE}=\sigma_{z}\otimes B_{z},
\end{align}
where $\sigma_{z}$ is the standard Pauli $Z$ operator and $B_{z}$ is an operator of the spin bath. To realize our one-qubit gates, we consider the following two orthonormal states
\begin{align}
&\ket{\nu_{1}(t)}=\cos\frac{\theta(t)}{2}\ket{0}+\sin\frac{\theta(t)}{2}e^{i\varphi(t)}\ket{1},
\notag\\
&\ket{\nu_{2}(t)}=\sin\frac{\theta(t)}{2}e^{-i\varphi(t)}\ket{0}-\cos\frac{\theta(t)}{2}\ket{1},
\end{align}
where $\theta(t)$ and $\varphi(t)$ are time-dependent parameters. By using the above two states and Eq. (\ref{hs}), one can get
\begin{align}
H_S=&-\frac{1}{2}\Big[\dot{\theta}(t)\sin\varphi(t)
+\dot{\varphi}(t)\sin\theta(t)\cos\theta(t)\cos\varphi(t)\Big]\sigma_{x}
\notag\\
&+\frac{1}{2}\Big[\dot{\theta}(t)\cos\varphi(t)
-\dot{\varphi}(t)\sin\theta(t)\cos\theta(t)\sin\varphi(t)\Big]\sigma_{y}
\notag\\
&+\frac{1}{2}\dot{\varphi}(t)\sin^{2}\theta(t)\sigma_{z},
\label{Gt1}
\end{align}
where $\sigma_{x}$, $\sigma_{y}$ and $\sigma_{z}$ are the standard Pauli operators acting on $\ket{0}$ and $\ket{1}$. If we keep the parameter $\varphi(t)$ being constant in each segment of the evolution, the Hamiltonian $H_S$ in Eq. (\ref{Gt1}) turns into a piecewise operator. In the $j$-th segment, the Hamiltonian $H_S$ is equal to
\begin{align}
H_{j}=\Omega_{j}(-\sin\varphi_{j}\sigma_{x}+\cos\varphi_{j}\sigma_{y})
\label{Gt2}
\end{align}
where $\Omega_{j}=\dot{\theta}(t)/2$. Based on $H_{SE}$, the unitary operator $V(t)$ can be chosen as
\begin{align}
V(t)=e^{-2\pi in\sigma_{x}t/\tau},\label{Vt1}
\end{align}
where $n$ is a positive integer. One can verify that the unitary operator $V(t)$ satisfies the two conditions in Eqs.~(\ref{condition1}) and (\ref{condition2}). Substituting Eqs.~(\ref{Gt2}) and (\ref{Vt1}) into Eq. (\ref{HS}), we can obtain the system Hamiltonian $H^\prime_{S}$. In the $j$-th segment, the system Hamiltonian $H^\prime_{S}$ reads
\begin{align}
H^\prime_{j}=&(n\omega-\Omega_{j}\sin\varphi_{j})\sigma_{x}+\Omega_{j}
\cos\varphi_{j}\cos(2n\omega t)\sigma_{y}\notag\\
&+\Omega_{j}\cos\varphi_{j}\sin(2n\omega t)\sigma_{z},
\label{HS1}
\end{align}
where $\omega={2\pi}/{\tau}$. The above Hamiltonian can be realized by adjusting the frequency, amplitude, and phase of the driving microwave field that couples the two Zeeman levels $\ket{m_{s}=0}$ and $\ket{m_{s}=-1}$.

To calculate the evolution operator generated by the above system Hamiltonian, we respectively consider the states $\ket{\phi_{1}(0)}$ and $\ket{\phi_{2}(0)}$ as the initial state of the system. The states $\ket{\phi_{1}(0)}$ and $\ket{\phi_{2}(0)}$ are defined by
\begin{align}
\ket{\phi_1(0)}&\equiv\ket{\nu_{1}(0)}=\cos\frac{\theta_0}{2}\ket{0}+\sin\frac{\theta_0}{2}e^{i\varphi_0}\ket{1},
\notag\\
\ket{\phi_{2}(0)}&\equiv\ket{\nu_{2}(0)}=\sin\frac{\theta_0}{2}e^{-i\varphi_0}\ket{0}-\cos\frac{\theta_0}{2}\ket{1},
\end{align}
where $\theta_0=\theta(0)$ and $\varphi_0=\varphi(0)$. One can find that the system evolves cyclically if the initial state of the system is $\ket{\phi_1(0)}$ or $\ket{\phi_2(0)}$. By using Eq. (\ref{UT}), one can get the evolution operator
\begin{align}
U(T)=e^{-i\gamma(T)}\ket{\phi_{1}(0)}\bra{\phi_{1}(0)}
+e^{i\gamma(T)}\ket{\phi_{2}(0)}\bra{\phi_{2}(0)},
\label{UT1}
\end{align}
where the phase
\begin{align}
\gamma(T)=\frac{1}{2}\int^{T}_{0}[1-\cos\theta(t)]\dot{\varphi}(t)dt.
\end{align}
The operator $U(T)$ in Eq. (\ref{UT1}) can also be written as
\begin{align}
U(T)=e^{-i\gamma(T)\boldsymbol{\mathrm{n}\cdot\sigma}},
\label{UT2}
\end{align}
where $\boldsymbol{\mathrm{n}}=(\sin\theta_{0}\cos\varphi_{0},
\sin\theta_{0}\sin\varphi_{0},\cos\theta_{0})$ is a unit vector and $\boldsymbol{\sigma}=(\sigma_{x},\sigma_{y},\sigma_{z})$. The above equation clearly shows that arbitrary one-qubit coherence-protected nonadiabatic geometric gates can be realized.

\subsection{Two-qubit gates}

In the above subsection, we have shown the realization of one-qubit coherence-protected nonadiabatic geometric gates. To realize universal quantum computation, one also needs an entangling two-qubit coherence-protected nonadiabatic geometric gate. In the following, we demonstrate how to realize a controlled coherence-protected nonadiabatic geometric gate.
\begin{figure}[t]
\begin{center}
\includegraphics[scale=0.275]{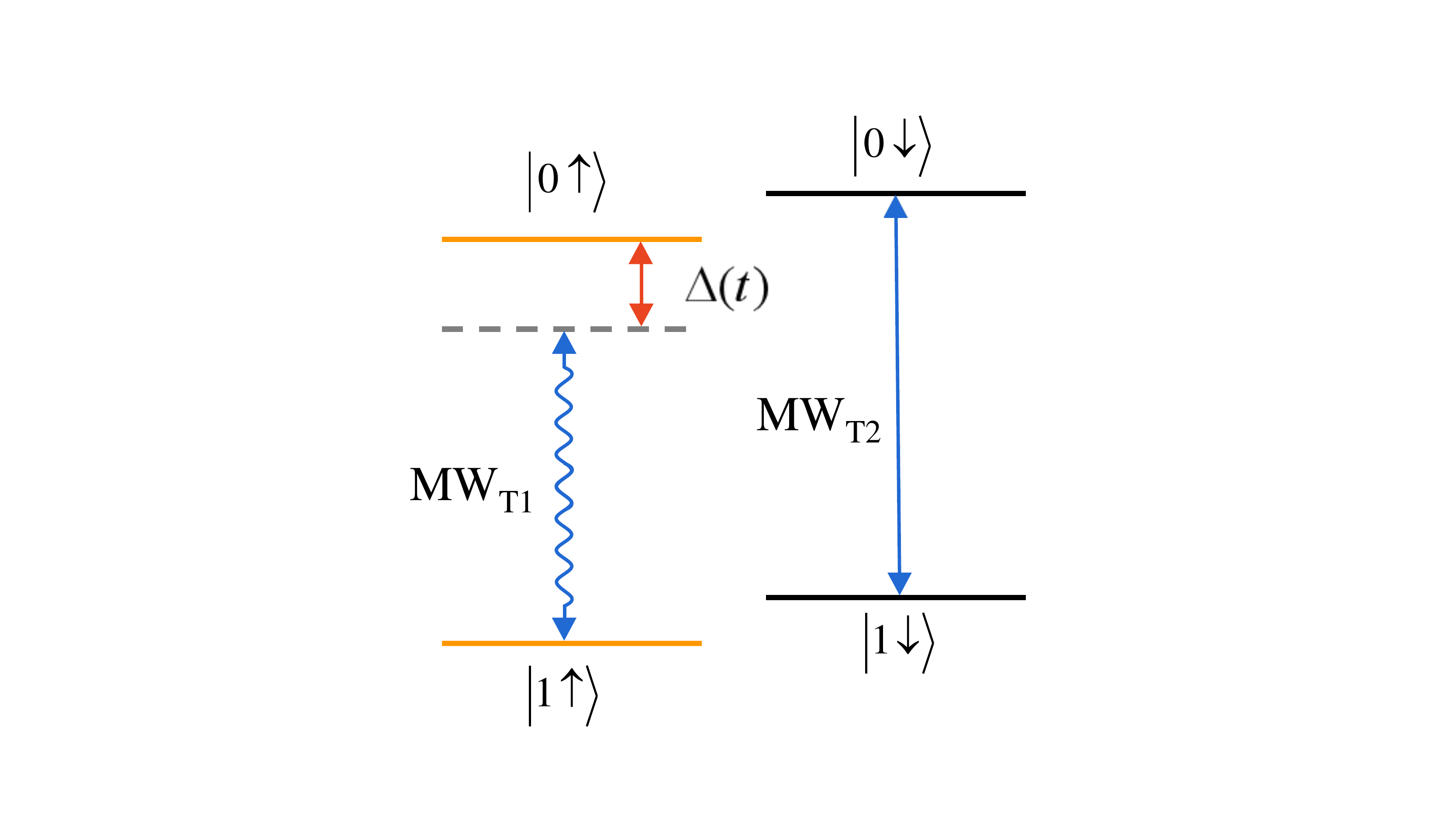}
\end{center}
\caption{(Color online) The level structure for two-qubit gates. The hyperfine interaction of the electron spin with the $^{13}\text{C}$ nuclear spin can be described by this four-level structure. The states $\ket{0,\uparrow}$ and $\ket{1,\uparrow}$ are coupled by microwave field MV$_{T1}$, with adjustable Rabi frequency, phase and detuning. The states $\ket{0,\downarrow}$ and $\ket{1,\downarrow}$ are coupled by microwave field MV$_{T2}$.
}\label{Fig2}
\end{figure}

We consider a NV center electron spin and one nearby $^{13}\text{C}$ nuclear spin. We respectively exploit the electron spin and nuclear spin as the target and control qubits. For the electron spin, we use the two Zeeman levels $\ket{m_{s}=-1}$ and $\ket{m_{s}=0}$ to represents the logical states $\ket{0}$ and $\ket{1}$ respectively, while for the nuclear spin, we directly use the basis states $\ket{\uparrow}$ and $\ket{\downarrow}$ as the computational states. The electron spin and nuclear spin are coupled to each other through hyperfine interaction, and the resultant levels are coupled by state-selective microwave and radio-frequency fields, as shown in Fig.~\ref{Fig2}. Because the gyromagnetic ratio of $^{13}\text{C}$ nuclear is about three orders of magnitude smaller than the electron gyromagnetic ratio \cite{nv6}, the electron spin is much more sensitive to the magnetic noise than the $^{13}\text{C}$ nuclear spin. For this reason, we only consider the affection of quasi-static magnetic noise, which is caused by the surrounding $^{13}\text{C}$ nuclear spin bath, on the electron spin, while ignore the affection of the noise on the $^{13}\text{C}$ nuclear spin. In this case, the system-environment interaction Hamiltonian reads
\begin{align}
H_{SE}=\sigma_{z}\otimes B_{z},
\end{align}
where $\sigma_{z}$ is the standard Pauli $Z$ operator acting on the electron spin and $B_{z}$ is an operator of its spin bath.
To realize our entangling gate, we consider the following four orthonormal states
\begin{align}
\ket{\nu_{1}(t)}=&\ket{0\downarrow},\notag\\
\ket{\nu_{2}(t)}=&\ket{1\downarrow},
\notag\\
\ket{\nu_{3}(t)}=&\cos\frac{\alpha(t)}{2}\ket{0\uparrow}+\sin\frac{\alpha(t)}{2}e^{i\beta(t)}\ket{1\uparrow},
\notag\\
\ket{\nu_{4}(t)}=&\sin\frac{\alpha(t)}{2}e^{-i\beta(t)}\ket{0\uparrow}-\cos\frac{\alpha(t)}{2}\ket{1\uparrow},
\end{align}
where $\alpha(t)$ and $\beta(t)$ are time-dependent parameters. It is interesting to know that $\ket{\nu_{3}(t)}$ and $\ket{\nu_{4}(t)}$ always reside inside the subspace spanned by ${\ket{0\uparrow}}$ and ${\ket{1\uparrow}}$. By using the above four states and Eq. (\ref{hs}), we can obtain $H_S$ and in the $j$-th segment, the Hamiltonian $H_S$ is
\begin{align}
H_{j}=\Omega_{j}(-\sin\beta_{j}\sigma_{x}+\cos\beta_{j}\sigma_{y})\otimes \ket{\uparrow}\bra{\uparrow}, \label{Gt3}
\end{align}
where $\Omega_{j}=\dot{\alpha}(t)/2$. Based on the interaction Hamiltonian $H_{SE}$, the unitary operator $V(t)$ for the entangling gate can be chosen as
\begin{align}
V(t)=e^{-2\pi in\sigma_{x}t/\tau}\otimes I_n,
\label{Vt2}
\end{align}
where $I_n$ is the identity operator acting on the nuclear spin. Substituting Eqs. (\ref{Gt3}) and (\ref{Vt2}) into Eq. (\ref{HS}), we obtain $H^\prime_{S}$ and in the $j$-th segment, it reads
\begin{align}
&H^\prime_{j}=[(n\omega-\Omega_{j}\sin\beta_{j})\sigma_{x}+\Omega_{j}\cos\beta_{j}\cos(2n\omega t)\sigma_{y}\notag\\
&+\Omega_{j}\cos\beta_{j}\sin(2n\omega t)\sigma_{z}]\otimes\ket{\uparrow}\bra{\uparrow}+n\omega\sigma_{x}
\otimes\ket{\downarrow}\bra{\downarrow}, \label{hGt3}
\end{align}
This Hamiltonian is experimentally feasible and can be realized by adjusting the frequency, amplitude, and phase of the state-selective microwave field. To calculate the evolution operator generated by the system Hamiltonian in the above equation, we respectively consider the states $\ket{\phi_k(0)}$ with $k\in\{1,2,3,4\}$ as the initial state of the system, where $\ket{\phi_k(0)}=\ket{\nu_{k}(0)}$. The system will evolve cyclically if the initial state of the system is one of $\ket{\phi_k(0)}$. With the help of $\ket{\phi_k(0)}$ and Eq. (\ref{UT}), we can get the corresponding evolution operator
\begin{align}
U(T)=\ket{\downarrow}\bra{\downarrow}\otimes I_e+\ket{\uparrow}\bra{\uparrow}\otimes e^{-i\gamma\boldsymbol{\mathrm{n}\cdot\sigma}},
\end{align}
where the phase $\gamma(T)=\int^{T}_{0}[1-\cos\alpha(t)]\dot{\beta}(t)/2 dt$ and $I_e$ is the identity operator acting on the electron spin. The gate $U(T)$ is a two-qubit controlled gate and takes the controlled-not and controlled-phase gates as special cases. Thus, an entangling coherence-protected nonadiabatic geometric gate has been realized. Combining this entangling gate with the one-qubit gates realized in subsection A, universal coherence-protected nonadiabatic geometric quantum computation can be implemented.

\section{Discussion}
\begin{figure}[t]
\begin{center}
\includegraphics[scale=0.375]{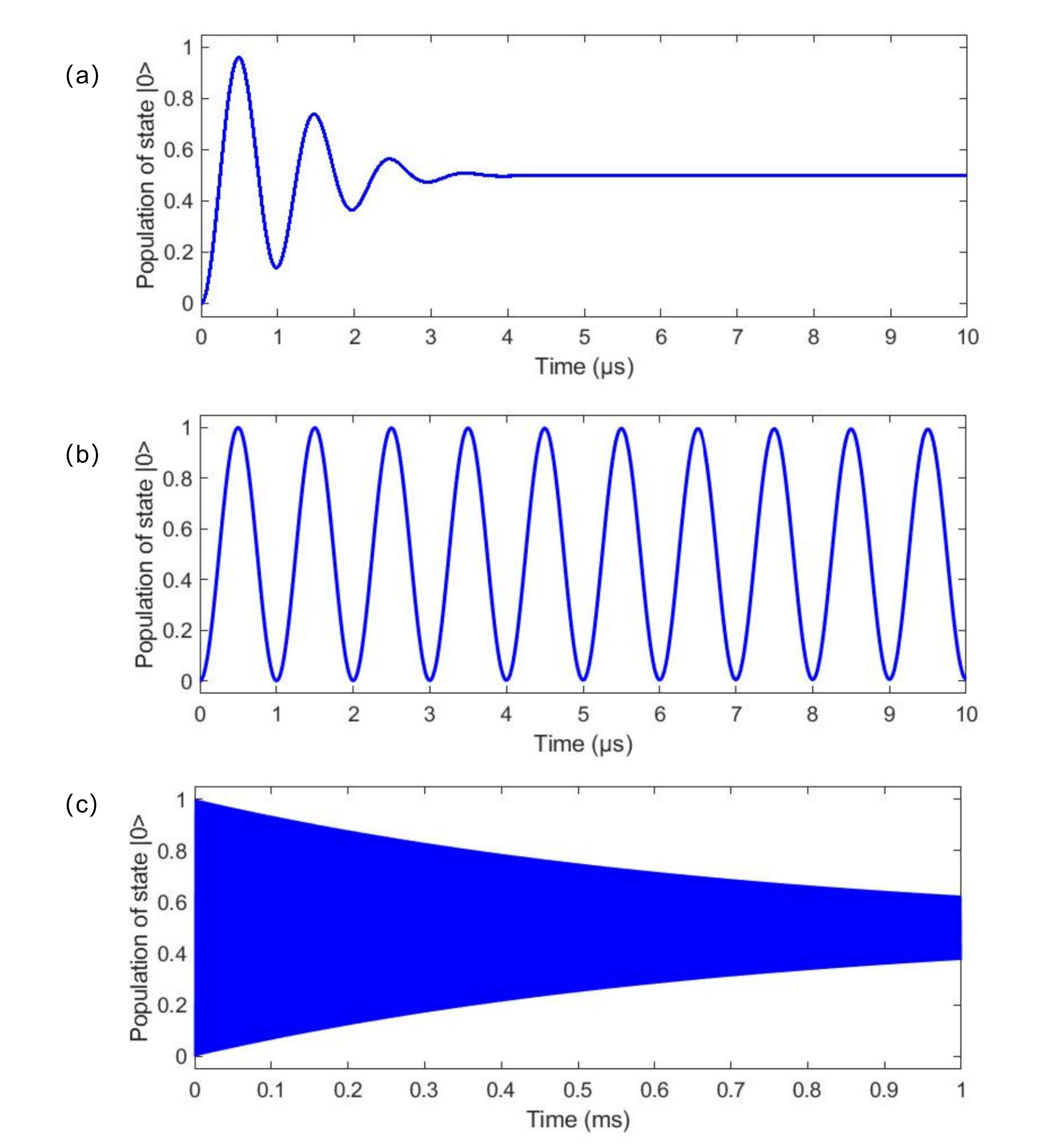}
\end{center}
\caption{(Color online) Free induction decay of the NV center electron spin coherence. (a) Unprotected FID process of the NV center electron spin coherence with parameters $\Delta=1$ MHz and $\sigma=\pi\times0.13$ MHz. (b) and (c) Protected FID process of the NV center electron spin coherence with parameters $\Delta=1$ MHz, $\sigma=\pi\times0.13$ MHz and $\tau=0.01$ $\mu$s.
}\label{Fig3}
\end{figure}
In this section, we simulate the practical performance of our scheme by using the quantum master equation
\begin{align}
\frac{d}{dt}\rho=-i[H(t),\rho]+\frac{\Gamma}{2}\sum_{a=\sigma_{\pm}}(2a^{\dag}\rho a-\rho a a^{\dag}-aa^{\dag}\rho),
\end{align}
where $H(t)$ is the system Hamiltonian, the Lindblad operators $a$, $a^{\dag}$ represent the spin relaxation process, and $\Gamma$ corresponds to the longitudinal spin relaxation time $T_{1}$ of the NV center electron spin. We here take $\Gamma=1$ KHz, which is proper for NV centers \cite{udd4}.

We first simulate the coherence time of the NV center electron spin qubit. We use free induction decay (FID) time to describe the transverse spin relaxation
time (dephasing time) $T_{2}$ of the NV center electron spin qubit \cite{nv7}. The evolution of the electron spin is governed by the system Hamiltonian $H_{\text{fid}}=\pi\Delta\sigma_{z}$, where $\Delta=1$ MHz. In practice, the system Hamiltonian turns into $H_{\text{fid}}^{\ast}=(\pi\Delta+\delta_{0})\sigma_{z}$. In the above, $\delta_{0}$ is an extra detuning which results from the Overhauser field. As the dynamical fluctuation of the Overhauser field is much slower than the typical gate time, it can be taken as a quasi-static random constant. We assume that $\delta_{0}$ satisfies a Gaussian distribution $P_{0}(\delta_{0})=\exp(-\delta_{0}^{2}/2\sigma^{2})/(\sigma\sqrt{2\pi})$, where $\sigma$ is the standard deviation of the distribution and we here take a standard deviation $\sigma=\pi\times0.13$ MHz \cite{nv7}.
Fig.~\ref{Fig3}(a) shows the unprotected FID process of the NV center electron spin. From Fig.~\ref{Fig3}(a), one can see the rapid free induction decay of the electron spin coherence, which is caused by the thermal distribution of the Overhauser field, and the decay time of FID is about $T_{2}=1.7$ $\mu $s. To prolong the FID time of the NV center electron spin qubit, we redesign the system Hamiltonian by using our scheme. The new system Hamiltonian corresponding to $H_{\text{fid}}$ is $H_{\text{fid}}^\prime=\pi\Delta(\cos(2\omega t)\sigma_{z}-\sin(2\omega t)\sigma_{y})+\omega\sigma_{x}$, where $\omega=2\pi/\tau$ with $\tau$ being set as $0.01$ $\mu$s. Fig.~\ref{Fig3}(b) and (c) show the protected FID process of the NV center electron spin by using our scheme. Note that Fig.~\ref{Fig3}(b) is one part of Fig~\ref{Fig3}(c). Specifically, Fig.~\ref{Fig3}(b) shows the case of time from $0$ $\mu$s to $10$ $\mu$s while Fig.~\ref{Fig3}(c) shows the case of time from $0$ ms to $1$ ms. From Fig.~\ref{Fig3}(b), one can see that in the typical gate time the oscillation of FID signal is well maintained and the residual effect of the Overhauser field noise result from higher order effect is negligible for the typical gate time.
From Fig.~\ref{Fig3}(c), one can see that under protection the $T_{2}$ of NV center electron spin can be extended up to the order of magnitude of about $1$ ms, which is close to the limit of $T_{1}$. This shows that our scheme is efficient for increasing the dephasing time $T_{2}$ of the NV center electron spin.

\begin{figure}[h]
\begin{center}
\includegraphics[scale=0.35]{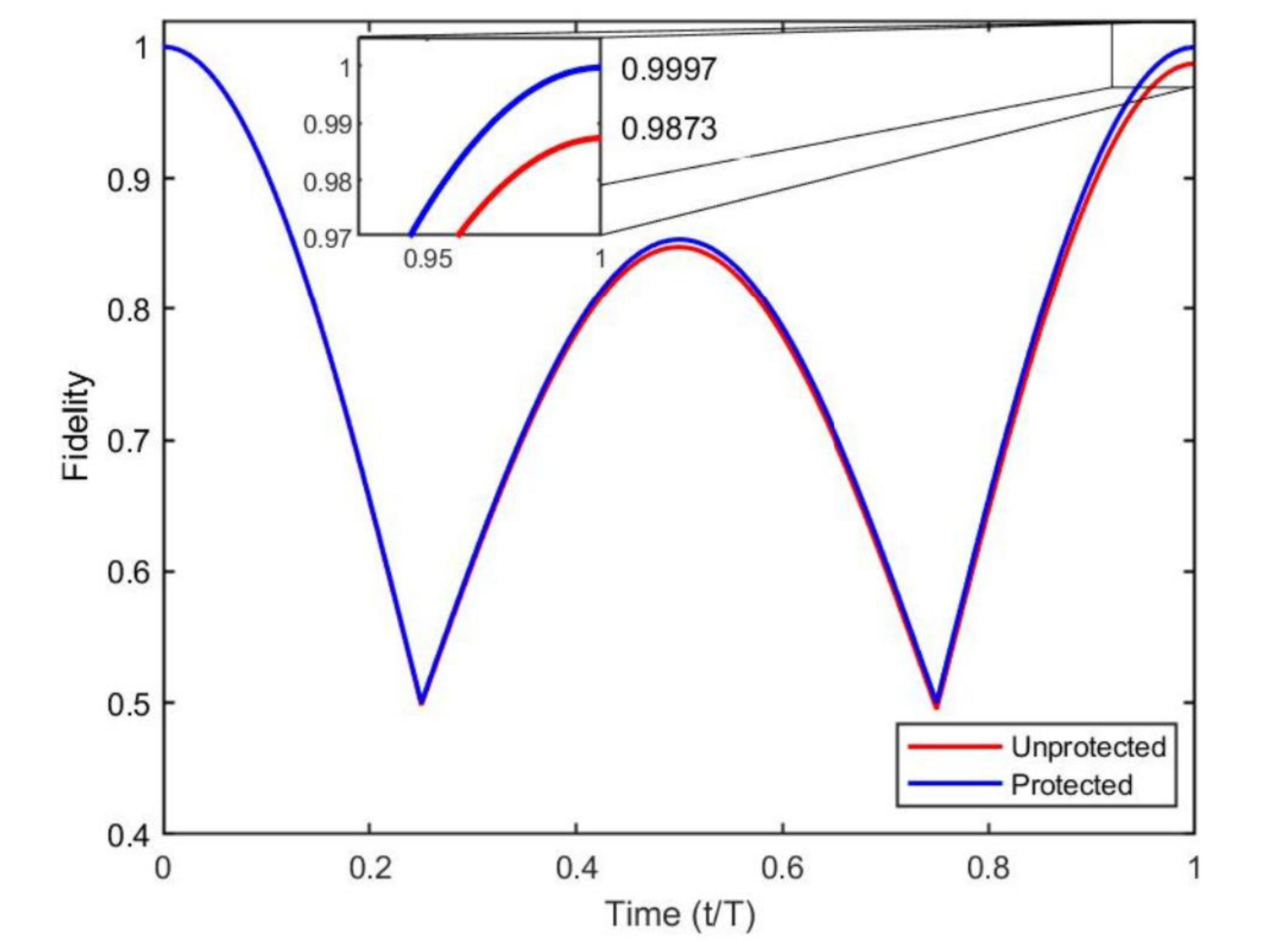}
\end{center}
\caption{(Color online) Fidelity dynamics as a function of $t/T$ for the gate $U(T)=\exp(-i\pi\sigma_{x}/4)$ with initial state $(\ket0+\ket1)/\sqrt{2}$ under the influence of the Overhauser field. The red curve shows the fidelity of the unprotected case. The blue curve shows the fidelity of the protected case with $\tau=0.0125$ $\mu$s.
 The fidelity can be up to $99.97\%$ under protection.
}\label{Fig4}
\end{figure}

We next simulate the performance of our coherence-protected nonadiabatic geometric gates under the influence of the Overhauser field. We take the gate $U(T)=\exp(-i\pi\sigma_{x}/4)$ as the test case. We set the Rabi frequency $\Omega_{j}$ in each segment as $2\pi$ MHz, the operation time $T$ of nonadiabatic geometric gate as $0.5$ $\mu$s. Our numerical result indicates that the fidelity is $98.73\%$ for the initial state $(\ket0+\ket1)/\sqrt{2}$ under the influence of the Overhauser field with standard deviation $\sigma=\pi\times0.13$ MHz, shown in Fig.~\ref{Fig4} by the red curve. For the protected case, we keep the parameters $\Omega_{j}=2\pi$ MHz and $\sigma=\pi\times0.13$ MHz unchanged and set $\tau=0.0125$ $\mu$s. One can see that the fidelity can be up to $99.97\%$ under the influence of the Overhauser field, which is shown in Fig.~\ref{Fig4} by the blue curve. This result shows that our scheme is efficient for increasing the fidelity of nonadiabatic geometric gates under the influence of quasi-static random noise.

\begin{figure}[h]
\begin{center}
\includegraphics[scale=0.35]{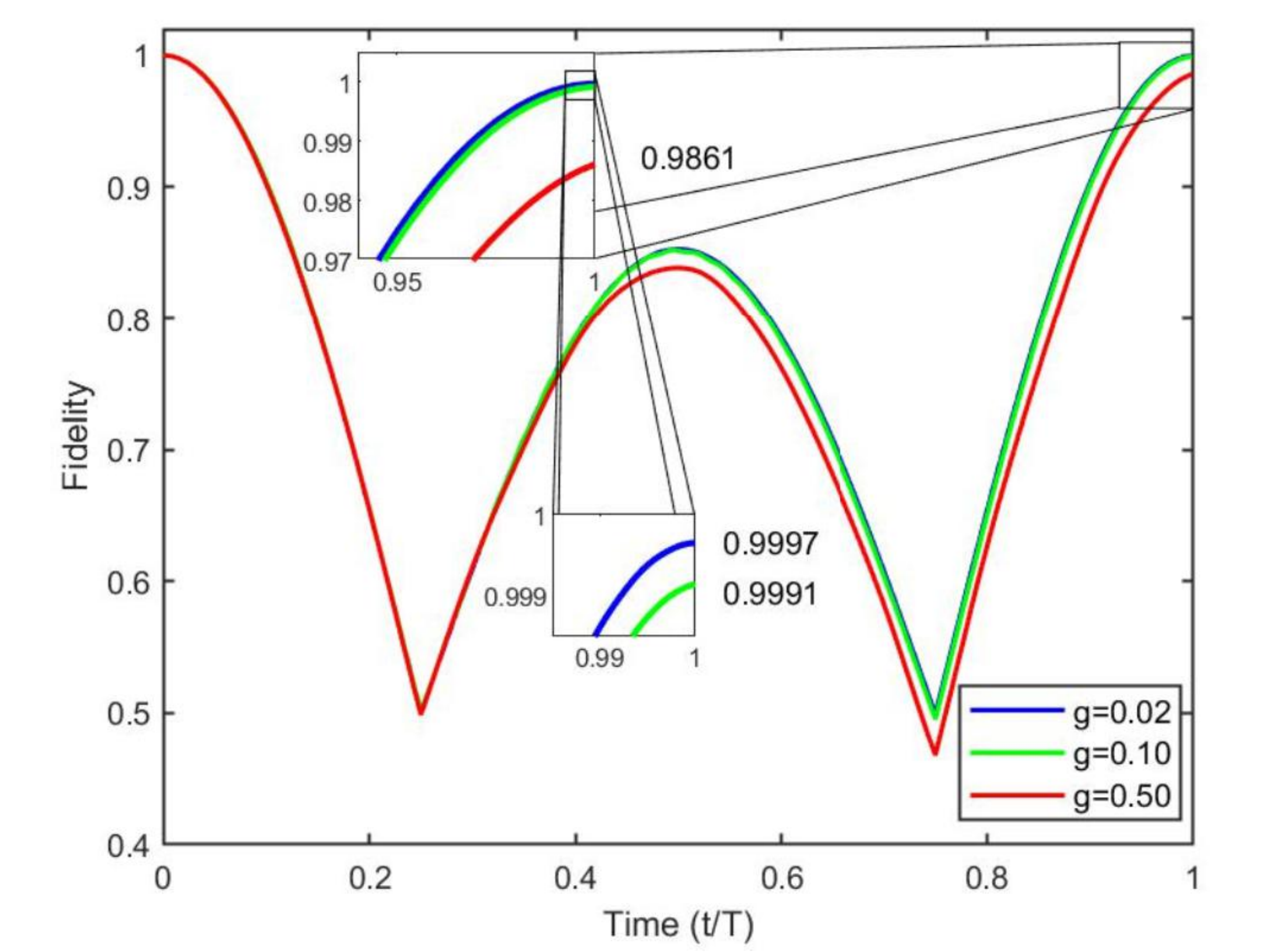}
\end{center}
\caption{(Color online) Fidelity dynamics as a function of $t/T$ for the gate $U(T)=\exp(-i\pi\sigma_{x}/4)$ with initial state $(\ket0+\ket1)/\sqrt{2}$ under the influence of time-dependent noise with correlation time $\tau_{e}=0.25$ $\mu$s. The blue curve shows the fidelity of the gate with $\tau=0.005$ $\mu$s and $g=0.02$. The green curve shows the fidelity of the gate with $\tau=0.025$ $\mu$s and $g=0.1$. The red curve shows the fidelity of the gate with $\tau=0.125$ $\mu$s and $g=0.5$.}
\label{Fig5}
\end{figure}
In the preceding discussion, since the dynamical
fluctuation of the Overhauser field in the NV center is much slower than the typical gate time, it is taken as a quasi-static random constant, which has an infinite correlation time. We now discuss the case of time-dependent noise with finite correlation time. We model the time-dependent interaction Hamiltonian $H_{SE}=\delta_{0}(t)\sigma_{z}\otimes B_{z}$ by Ornstein-Uhlenbeck processes \cite{udd4,ou1,ou2}, where $\delta_{0}(t)$ is a random variable obtained by solving the Ornstein-Uhlenbeck equation
\begin{align}
d\delta_{0}=-\frac{\delta_{0}-\mu}{\tau_{e}}dt+\sigma\sqrt{\frac{2}{\tau_{e}}}dW.
\end{align}
In the above, $\mu$ is the mean, $\sigma$ is the standard deviation, $\tau_{e}$ is the correlation time, and $W$ is the standard Wiener process. Here we set $\tau_{e}=0.25$ $\mu$s, $\sigma=1.5$ MHz, the mean $\mu$ as zero. We simulate the performance of the gate $U(T)=\exp(-i\pi\sigma_{x}/4)$ with different $g=\tau/\tau_{e}$, and the result is shown in Fig.~\ref{Fig5}. One can see that when $\tau=0.005$ $\mu$s and $g=0.02$, the dynamical decoupling condition $\tau/\tau_{e}\ll1$ is well fulfilled and the fidelity is up to $99.97\%$. For the case of $\tau=0.025$ $\mu$s and $g=0.1$, the dynamical decoupling condition is fulfilled barely and the fidelity is down to $99.91\%$. When $\tau=0.125$ $\mu$s and $g=0.5$, the dynamical decoupling condition is not fulfilled and the fidelity is down to $98.61\%$. The above results indicate that as long as $\tau$ is short enough and the dynamical decoupling condition is fulfilled, our coherence-protected nonadiabatic geometric gates can be resilient to time-dependent noise efficiently.

\section{Conclusion}

In conclusion, we have demonstrated the realization of coherence-protected nonadiabatic geometric gates. Our result shows that based on the system Hamiltonian $H_{S}$ that realizes a nonadiabatic geometric gate, one may construct a new system Hamiltonian $H^\prime_{S}$ with the help of the dress operator $V(t)$. The new system Hamiltonian $H^\prime_{S}$ allows one to preserve the geometric robustness while decoupling the system from its environment, resulting in coherence-protected nonadiabatic geometric gates. Our scheme does not need extra resources like auxiliary systems or the encoding of logical qubits with physical qubits. Thus it is experimentally friendly. To show the application of our scheme, we give the implementation of our scheme with NV centers and show the specific realization procedure of a universal set of coherence-protected nonadiabatic geometric gates. We hope our scheme can shed light on the realization of more efficient quantum gates.

\begin{acknowledgments}
The authors acknowledge support from the National Natural Science Foundation of China through Grant No. 11775129.
\end{acknowledgments}


\begin{thebibliography}{99}
\bibitem{Shor} P. W. Shor, SIAM J. Comput. \textbf{26}, 1484 (1997).
\bibitem{Grover} L. K. Grover, Phys. Rev. Lett. \textbf{79}, 325 (1997).
\bibitem{Chiara} G. De Chiara and G. M. Palma, Phys. Rev. Lett. \textbf{91}, 090404 (2003).
\bibitem{Solinas2004} P. Solinas, P. Zanardi, and N. Zangh\`{i}, Phys. Rev. A \textbf{70}, 042316 (2004).
\bibitem{Zhu2005} S. L. Zhu, Z. D. Wang, and P. Zanardi, Phys. Rev. Lett. \textbf{94}, 100502 (2005).
\bibitem{Lupo} C. Lupo, P. Aniello, M. Napolitano, and G. Florio, Phys. Rev. A \textbf{76}, 012309 (2007).
\bibitem{Filipp} S. Filipp, J. Klepp, Y. Hasegawa, C. Plonka-Spehr, U. Schmidt, P. Geltenbort, and H. Rauch, Phys. Rev. Lett. \textbf{102}, 030404 (2009).
\bibitem{Johansson} M. Johansson, E. Sj\"{o}qvist, L. M. Andersson, M. Ericsson, B. Hessmo, K. Singh, and D. M. Tong, Phys. Rev. A \textbf{86}, 062322 (2012).
\bibitem{Berger} S. Berger, M. Pechal, A. A. Abdumalikov, Jr., C. Eichler, L. Steffen, A. Fedorov, A. Wallraff, and S. Filipp, Phys. Rev. A \textbf{87}, 060303(R) (2013).
\bibitem{Jones} J. A. Jones, V. Vedral, A. Ekert, and G. Castagnoli, Nature \textbf{403}, 869 (2000).
\bibitem{Berry} M. V. Berry, Proc. R. Soc. Lond. A \textbf{392}, 45 (1984).
\bibitem{Duan} L. M. Duan, J. I. Cirac, and P. Zoller, Science \textbf{292}, 1695 (2001).
\bibitem{Zanardi} P. Zanardi and M. Rasetti, Phys. Lett. A \textbf{264}, 94 (1999).
\bibitem{Wilczek} F. Wilczek and A. Zee, Phys. Rev. Lett. \textbf{52}, 2111 (1984).
\bibitem{WangXB} W. Xiang-Bin and M. Keiji, Phys. Rev. Lett. \textbf{87}, 097901 (2001).
\bibitem{Zhu} S. L. Zhu and Z. D. Wang, Phys. Rev. Lett. \textbf{89}, 097902 (2002).
\bibitem{Aharonov} Y. Aharonov and J. Anandan, Phys. Rev. Lett. \textbf{58}, 1593 (1987).
\bibitem{Sjoqvist2012} E. Sj\"oqvist, D. M. Tong, L. M. Andersson, B. Hessmo, M. Johansson, and K. Singh, New J. Phys. \textbf{14}, 103035 (2012).
\bibitem{Xu2012} G. F. Xu, J. Zhang, D. M. Tong, E. Sj\"oqvist, and L. C. Kwek, Phys. Rev. Lett. \textbf{109}, 170501 (2012).
\bibitem{Anandan} J. Anandan, Phys. Lett. A \textbf{133}, 171 (1988).
\bibitem{WANG2001} X. Wang and K. Matsumoto, J. Phys. A: Math. Gen. \textbf{34}, L631 (2001).
\bibitem{Zhu2003} S. L. Zhu and Z. D. Wang, Phys. Rev. Lett. \textbf{91}, 187902 (2003).
\bibitem{Zhu2003PRA} S. L. Zhu and Z. D. Wang, Phys. Rev. A \textbf{67}, 022319 (2003).
\bibitem{Sjoqvist2003} A. Friedenauer and E. Sj\"oqvist, Phys. Rev. A \textbf{67}, 024303 (2003).
\bibitem{Solinas} P. Solinas, P. Zanardi, N. Zanghi, and F. Rossi, Phys. Rev. A \textbf{67}, 052309 (2003).
\bibitem{Zheng} S. B. Zheng, Phys. Rev. A \textbf{70}, 052320 (2004).
\bibitem{Zhang2005} X. D. Zhang, S. L. Zhu, L. Hu, and Z. D. Wang, Phys. Rev. A \textbf{71}, 014302 (2005).
\bibitem{Chen2006} C. Y. Chen, M. Feng, X. L. Zhang, and K. L. Gao, Phys. Rev. A \textbf{73}, 032344 (2006).
\bibitem{Cen2006} L. X. Cen, Z. D. Wang, and S. J. Wang, Phys. Rev. A \textbf{74}, 032321 (2006).
\bibitem{Feng2007} X. L. Feng, Z. S. Wang, C. F. Wu, L. C. Kwek, C. H. Lai, and C. H. Oh, Phys. Rev. A \textbf{75}, 052312 (2007).
\bibitem{Wu2007} C. F. Wu, Z. S. Wang, X. L. Feng, H. S. Goan, L. C. Kwek, C. H. Lai, and C. H. Oh, Phys. Rev. A \textbf{76}, 024302 (2007).
\bibitem{Kim} K. Kim, C. F. Roos, L. Aolita, H. H\"{a}ffner, V. Nebendahl, and R. Blatt, Phys. Rev. A  \textbf{77}, 050303(R) (2008).
\bibitem{Feng2009} X. L. Feng, C. F. Wu, H. Sun, and C. H. Oh, Phys. Rev. Lett. \textbf{103}, 200501 (2009).
\bibitem{Ota} Y. Ota, Y. Goto, Y. Kondo, and M. Nakahara, Phys. Rev. A \textbf{80}, 052311 (2009).
\bibitem{Oto2009} Y. Ota and Y. Kondo, Phys. Rev. A \textbf{80}, 024302 (2009).
\bibitem{Thomas2011} J. T. Thomas, M. Lababidi, and M. Tian, Phys. Rev. A \textbf{84}, 042335 (2011).
\bibitem{Chen2012} Y. Y. Chen, X. L. Feng, and C. H. Oh, Opt. Commun. \textbf{285}, 5554 (2012).
\bibitem{Xu2014SR} G. F. Xu and G. L. Long, Sci. Rep. \textbf{4}, 6814 (2014).
\bibitem{Xu2014PRA} G. F. Xu and G. L. Long, Phys. Rev. A \textbf{90}, 022323 (2014).
\bibitem{Zhao2016} P. Z. Zhao, G. F. Xu, and D. M. Tong, Phys. Rev. A \textbf{94}, 062327 (2016).
\bibitem{Zhu2016} Z. T. Liang, X. Yue, Q. Lv, Y. X. Du, W. Huang, H. Yan, and S. L. Zhu, Phys. Rev. A \textbf{93}, 040305(R) (2016).
\bibitem{Zhao} P. Z. Zhao, X. D. Cui, G. F. Xu, E. Sj\"{o}qvist, and D. M. Tong, Phys. Rev. A \textbf{96}, 052316 (2017).
\bibitem{Xue} T. Chen and Z. Y. Xue, Phys. Rev. Appl. \textbf{10}, 054051 (2018).
\bibitem{Liu} B. J. Liu, X. K. Song, Z. Y. Xue, X. Wang, and M. H. Yung, Phys. Rev. Lett. \textbf{123}, 100501 (2019).
\bibitem{xia2020} Y. H. Kang, Z. C. Shi, B. H. Huang, J. Song, and Y. Xia, Phys. Rev. A \textbf{101}, 032322 (2020).
\bibitem{xue2020} C. Zhang, T. Chen, S. Li, X. Wang, and Z. Y. Xue, Phys. Rev. A \textbf{101}, 052302 (2020).
\bibitem{own2020} K. Z. Li, P. Z. Zhao, and D. M. Tong, Phys. Rev. Res. \textbf{2}, 023295
(2020).
\bibitem{su2020} C. Y. Guo, L. L. Yan, S. Zhang, S. L. Su, and W. Li, Phys. Rev. A \textbf{102}, 042607 (2020).
\bibitem{Leibfried} D. Leibfried, B. De Marco, V. Meyer, D. Lucas, M. Barrett, J. Britton, W. M. Itano, B. Jelenkovi\'{c}, C. Langer, T. Rosenband, and D. J. Wineland, Nature \textbf{422}, 412 (2003).
\bibitem{Du} J. F. Du, P. Zou, and Z. D. Wang, Phys. Rev. A \textbf{74}, 020302(R) (2006).
\bibitem{npj2018}F. Kleissler, A. Lazariev, and S. Arroyo-Camejo, npj Quantum. Inf. \textbf{4}, 49 (2018).
\bibitem{guo2018} T. Wang, Z. Zhang, L. Xiang, Z. Jia, P. Duan, W. Cai, Z. Gong, Z. Zong, M. Wu, J. Wu, L. Sun, Y. Yin, and G. Guo, New J. Phys. \textbf{20}, 065003 (2018).
\bibitem{Sun} Y. Xu, Z. Hua, T. Chen, X. Pan, X. Li, J. Han, W. Cai, Y. Ma, H. Wang, Y. P. Song, Z. Y. Xue, and L. Sun, Phys. Rev. Lett. \textbf{124}, 230503 (2020).
\bibitem{Yin} P. Z. Zhao, Z. J. Z. Dong, Z. X. Zhang, G. P. Guo, D. M. Tong, and Y. Yin, arXiv:1909.09970.
\bibitem{udd1} K. M. Fonseca-Romero, S. Kohler, and P. Hnggi, Chem. Phys. \textbf{296}, 307 (2004).
\bibitem{udd2} F. F. Fanchini, J. E. M. Hornos, and R. d. J. Napolitano, Phys. Rev. A \textbf{75}, 022329 (2007).
\bibitem{udd3} P. Rabl, P. Cappellaro, M. V. Gurudev Dutt, L. Jiang, J. R. Maze, and M. D. Lukin, Phys. Rev. B \textbf{79}, 041302(R) (2009).
\bibitem{udd4} J. Cai, B. Naydenov, R. Pfeiffer, L. P. McGuinness, K. D. Jahnke, F. Jelezko, M. B. Plenio, and A. Retzker, New J. Phys. \textbf{14}, 113023 (2012).
\bibitem{udd5} A. Z. Chaudhry and J. B. Gong, Phys. Rev. A \textbf{85}, 012315 (2012).
\bibitem{udd6} A. Bermudez, F. Jelezko, M. B. Plenio, and A. Retzker, Phys. Rev. Lett. \textbf{107}, 150503 (2011).
\bibitem{ref10}V. Jacques, P. Neumann, J. Beck, M. Markham, D. Twitchen,J. Meijer, F. Kaiser, G. Balasubramanian, F. Jelezko, and J. Wrachtrup, Phys. Rev. Lett. \textbf{102}, 057403 (2009).
\bibitem{nv1} P. Huang, X. Kong, N. Zhao, F. Shi, P. Wang, X. Rong, R. B. Liu, and J. F. Du, Nature Commun. \textbf{2}, 570 (2011).
\bibitem{nv2} N. Zhao, S. W. Ho, and R. B. Liu, Phys. Rev. B \textbf{85}, 115303 (2012).
\bibitem{nv3} X. Rong, J. Geng, Z. Wang, Q. Zhang, C. Ju, F. Shi, C. K. Duan, and J. Du, Phys. Rev. Lett. \textbf{112}, 050503 (2014).
\bibitem{nv4} X. Wang, L. S. Bishop, J. P. Kestner, E. Barnes, K. Sun, and S. Das Sarma, Nat. Commun. \textbf{3}, 997 (2012).
\bibitem{nv5} H. De Raedt, B. Barbara, S. Miyashita, K. Michielsen, S. Bertaina, and S. Gambarelli, Phys. Rev. B \textbf{85}, 014408 (2012).
\bibitem{nv6} S. J. Peng, X. K. Xu, K. B. Xu, P. Huang, P. F. Wang, X. Kong, X. Rong, F. Z. Shi, C. K. Duan, and J. F. Du, Sci. Bull. \textbf{63}, 336 (2018).
\bibitem{nv7} X. Rong, J. Geng, F. Shi, Y. Liu, K. Xu, W. Ma, F. Kong, Z. Jiang, Y. Wu, and J. Du, Nat. Commun. \textbf{6}, 8748 (2015).

\bibitem{ou1}  S. Austin, M. Q.Khan, M. Mudassar, and A. Z. Chaudhry, Phys. Rev. A \textbf{100}, 022102 (2019).




\bibitem{ou2} K. Jacobs, \emph{Stochastic Processes for Physicists: Understanding Noisy Systems} (Cambridge University Press, Cambridge, 2010).


\end{thebibliography}
\end{document}